\documentclass{emulateapj}
\usepackage{natbib}
\newcommand{\Msun}{ M_{\odot}}
\newcommand{\Mdot}{\dot{M}}
\newcommand{\Mbh}{M_{\rm BH}}

\begin{document}
\title{A Comment on the Radiative Efficiency of AGN}
\author{Ari Laor\altaffilmark{1} and Shane Davis\altaffilmark{2}}
\altaffiltext{1}{Physics Department, Technion, Haifa~32000, Israel}
\altaffiltext{2}{Canadian Institute for Theoretical Astrophysics. Toronto, ON M5S3H4, Canada}

\begin{abstract}

A recent study of the accretion efficiency of the PG sample AGN, based
on the thin accretion disk emission model, finds the accretion
radiative efficiency is correlated with the black hole mass ($\eta
\propto M_{\rm BH}^{0.5}$).  A followup study suggests the correlation
is an artifact induced by selection effects.  Here we point out there
are two independent effects. The first is a sample selection effect,
which leads to a high $L/L_{\rm Edd}$ AGN sample.  The second effect is
the observed small spread in the SED shape, which is not induced by
selection effects. The second effect is what leads to the $\eta
\propto M_{\rm BH}^{0.5}$ relation in the PG sample. The physical reason for the small
spread in the SED shape is an intriguing open question.

\end{abstract}

\section*{}
Thin accretion disk emission models allow to measure the absolute
accretion rate in AGN, based on the optical luminosity $L_{\rm opt}$, 
if the black hole mass $\Mbh$
is known. For example, Davis \& Laor (2011, hereafter DL) derive
\begin{equation}
\Mdot = 3.5 \Msun~{\rm yr}^{-1} \left( L_{\rm opt,45}\right)^{3/2}
M_8^{-0.89},
\end{equation}
where $M_8=\Mbh/10^8 \Msun$, $L_{\rm opt} \equiv \nu L_\nu$ at 4686 \AA~  
and $L_{\rm opt,45}=L_{\rm opt}/{\rm  10^{45} \; erg \; s^{-1}}$. 
The radiative efficiency, $\eta$, is then 
\begin{equation}
\eta=\frac{L_{\rm bol}}{\Mdot c^2 }.
\end{equation}
where $L_{\rm bol}$ is the bolometric luminosity. DL applied this method to the
complete and well defined PG quasar sample, using 80 of the
87 $z<0.5$ AGN from the sample with UV observation. Single epoch
broad line region based estimates of $\Mbh$ are available for all objects. The results 
shows that $\eta$ tends to rise with $\Mbh$. A best fit relation gives
\begin{equation}
\eta=0.089M_8^{0.52} .
\end{equation}

The PG quasar sample is selected based on color, presence of broad permitted lines,
and point-like morphology. The morphology criterion implies
$L_{\rm opt}/L_{\rm host}>1$. In addition, the following relations hold, 
$L_{\rm host}\propto M_{\rm host}$,
$M_{\rm host}\ge M_{\rm bulge}$, where $M_{\rm host}$ and $M_{\rm bulge}$ 
are the host mass and bulge mass, and also $M_{\rm bulge}\propto \Mbh$. Thus, 
the morphology criterion
implies a lower limit on $L_{\rm opt}/\Mbh$, and therefore a lower limit on 
$L_{\rm bol}/L_{\rm Edd}$, if $L_{\rm bol}\propto L_{\rm opt}$. Indeed,
most PG quasars have $L_{\rm bol}/L_{\rm Edd}>0.1$ (e.g. Fig.13 in DL). In contrast 
with some other AGN surveys,
which can be significantly less contaminated by the host emission, like X-ray surveys, 
and therefore extend to much lower $L_{\rm bol}/L_{\rm Edd}$ values.

Raimundo et al. (2011, hereafter R11) used the thin accretion disk
method to derive $\eta$ for a small sample of AGN, part of which are
nearby Seyfert galaxies. These objects have a distribution of $\Mbh$
values similar to the PG sample, but $L_{\rm opt}$ typically an order
of magnitude lower (R11, Fig.17). The implied $\Mdot$ is a factor of
$\sim 30$ lower than in the PG sample (R11, Fig.12), as expected from
equation 1 above.  The implied typical $\eta$ values are $\sim 3$
times higher than for the PG quasars at similar $\Mbh$ (R11, Fig.16),
as expected from equation 2 above, if $L_{\rm opt}/L_{\rm bol}$ has the
same characteristic ratio as in the PG sample. R11 also analyzed a
handful of higher $z$ SDSS quasars, similar in properties to the PG
quasars, and derive $\eta$ values similar to the PGs for
similar $\Mbh$ values (R11, Fig.16). R11 study carefully the range of parameters
covered by the PG sample, assuming a fixed ratio for $L_{\rm
  opt}/L_{\rm bol}$, and conclude that the $\eta$ vs. $\Mbh$ relation
found for the PG quasar sample is an artifact of the sample selection
criteria.

Below we explain why the $\eta$ vs. $\Mbh$ relation in the PG sample
is not an artifact.  There are two independent relations among the
observables that generate the correlation, and each needs to be
carefully understood. The first comes from the PG sample selection
criterion. As explained above, this leads to a lower limit on $L_{\rm
  opt}/L_{\rm Edd}$, and thus to a lower limit on $\Mdot$ for a given
$\Mbh$. But, this selection effect alone does not produce an $\eta$
vs. $\Mbh$ relation. The second, and essential relation is the observed
small spread in $L_{\rm opt}/L_{\rm bol}$ (DL, Fig.12). Applying both
relations, $L_{\rm opt}\propto \Mbh$ and $L_{\rm opt}\propto L_{\rm
  bol}$ in eqs.1 \& 2 above, leads inevitably to a rise of $\eta$ with
$\Mbh$ (with a power of 0.39, in the absence of scatter). 
The observed correlation between $\eta$ and $\Mbh$ is only an artifact
if selection effects generate the $L_{\rm opt}\propto L_{\rm bol}$ relation.

Therefore, it is crucial to understand that the $L_{\rm opt}\propto L_{\rm
  bol}$ relation is not a selection effect of the PG sample.  The
value of $L_{\rm bol}$ is mostly set by the far UV to soft X-ray part
of the SED, which is independent of $L_{\rm opt}$. The value of
$L_{\rm opt}$ should come predominantly from larger disk radii than
the UV and X-rays, and is not expected to provide a measure of $L_{\rm
  bol}$. The expected SEDs of the PG sample, if all had a fixed $\eta$
($\sim 0.1$), are shown in DL Fig.7. There is no selection effect which
prevents the PG quasars from showing the predicted fixed $\eta$
SED. Yet, at low $\Mbh$ the observed SED is systematically colder than
the expected fixed $\eta$ SED, while at the highest $\Mbh$ the observed SED is
systematically hotter. The $\eta=0.1$ SED is expected to peak at $\sim 10$
Rydberg for the $\Mbh<10^7\Msun$ PG sample objects, and at $< 1$ Rydberg
for the $\Mbh>10^9\Msun$ objects. But, the SED generally peaks at 1 to
a few Rydberg, regardless of $\Mbh$.

Low $\Mbh$ AGN should have hotter accretion disks than high $\Mbh$
AGN. But they don't.  Why is this so?  If $\eta$ is
set by the black hole spin $a$, then the rise of $\eta$ with $\Mbh$
implies a rise of $a$ with $\Mbh$. In that case the innermost disk
radius $r_{\rm in}$, in units of $Rg=G\Mbh/c^2$, shrinks with
increasing $\Mbh$, compensating for the drop in the disk temperature
at a given radius (in $Rg$ units).  However, it is not clear why the two independent
effects, the drop in $r_{\rm in}$, and the rise in $\Mbh$,
happen to cancel out and produce a relatively uniform SED (as far as one can tell).

Another option is that the accretion disk structure differs from the
assumption of the thin disk model, due to some unmodelled physical
processes, which produce a typical characteristic SED. In that case, 
$\eta$ cannot be
used to derive $a$. However, the inferred $\eta$ relation may still
provide some useful clue to processes in the innermost part of the
accretion disk.

{}


\begin{thebibliography}{99}

\bibitem[Davis 
\& Laor(2011)]{2011ApJ...728...98D} Davis, S.~W., \& Laor, A.\ 2011, ApJ, 728, 98 

\bibitem[Raimundo et al.(2011)]{2011arXiv1109.6225R} Raimundo, S.~I., 
Fabian, A.~C., Vasudevan, R.~V., Gandhi, P., 
\& Wu, J.\ 2011, arXiv:1109.6225 

\end{thebibliography}
\end{document}